\def\mt {m_t}

\def\mH {m_H}
\def\bbar {b\bar b}
\def\lnu {\ell\nu}
\def\simlt {\stackrel{<}{\sim}}
\def\gev{\hbox{GeV}}
\documentstyle[12pt,world_sci]{article}
\begin{document}
\begin{flushright}
MSUHEP-40815
\end{flushright}
\title{{\bf DETECTING THE INTERMEDIATE-MASS NEUTRAL \\
HIGGS BOSON AT THE LHC THROUGH $pp \to W H$}}
\author{PANKAJ AGRAWAL and DAVID BOWSER-CHAO.\thanks{{\bf Presented at
the Eighth Meeting of the Division of Particles and Fields
Albuquerque, New Mexico, August 2-6, 1994.}}\thanks
{Talk presented by David Bowser-Chao.} \\
{\em Department of Physics \& Astronomy, Michigan State University \\
East Lansing, Michigan 48824, USA} \\
\vspace{0.3cm}
KINGMAN CHEUNG \\
{\em Department of Physics \& Astronomy, Northwestern University \\
Evanston, Illinois 60208, USA} \\
\vspace{0.3cm}
and \\
\vspace{0.3cm}
DUANE A. DICUS \\
{\em Center for Particle Physics, University of Texas at Austin \\
Austin, Texas 78712, USA}}
\maketitle
\setlength{\baselineskip}{2.6ex}

\begin{center}
\parbox{13.0cm}
{\begin{center} ABSTRACT \end{center}
{\small \hspace*{0.3cm}
We examine the exclusive signature $pp \to WH \to \ell \nu\;b
\bar b$ at the LHC. Although the backgrounds, principally arising from
top production and $Wjj$, are quite severe, it is shown that judicious
application of phase-space cuts and the use of $b$-tagging can in fact
greatly enhance the detectability of this channel.

}}
\end{center}

\section{Introduction}
Discovery of the intermediate mass Higgs boson is covered by

LEP II from below (through $e^+e^-\to ZH$), and from above by the

LHC (through $pp\to H\to ZZ^*$). The LHC strategy for the intermediate region
($100 \simlt m_H \simlt 120\,\gev$) is to use the the rare decay mode
$H\to\gamma\gamma$ along with very high di-photon mass resolution to
beat down both the very large irreducible background as well as the
even larger reducible background $pp\to\gamma+\hbox{jet}$. The
signal/background ratios are quite low, so a thorough understanding of
the systematic error in mass-resolution and jet/$\gamma$ rejection is
crucial.

 As an alternative/complement to this search mode, we consider

the primary decay mode of the Higgs boson, $H\to \bbar$ produced in the process
$pp\to WH \to \lnu\bbar$. Even with tagging of the lepton, however, the QCD
background $pp\to Wjj$ completely overwhelms the signal. With an efficient
means of tagging $b$-quark jets and effectively rejecting light quark
and gluon jets, this background could be cut down to the level of the
much smaller subprocess $pp\to W\bbar$; after Higgs boson mass
reconstruction, the latter is comparable to the signal\cite{agrawal}.

A potentially more serious threat comes from the expected copious
production of the top quark, whose current lower mass limit\cite{d0
limit} ($m_t > 131 \,\hbox{GeV} > m_b + m_W$)
implies a large source of both $b$-quarks and leptonically decaying
$W$ bosons. For this reason, we consider the {\it exclusive}
production of $pp\to WH$ --- by rejecting ``extra'' jets and leptons,
we can greatly suppress backgrounds from single and pair production of
top quarks\cite{stange}\footnote{ This reference examines the same
process considered here, but at the Tevatron (and its proposed upgrades
in luminosity and/or energy) and at the LHC, assuming Tevatron-type
acceptance and resolution.}.

\section{Backgrounds}
The backgrounds considered include:
\begin{eqnarray}
Wjj      &\to& \lnu jj \;, \label{eq:wjj}\\

Wb\bar b        &\to& \lnu \bbar \;, \label{eq:wbb}\\

WZ              &\to& \lnu \bbar \;, \label{eq:wz}\\

t\bar t          &\to& (\lnu\lnu,\lnu jj)+\bbar \;, \label{eq:tt} \\

tbq            &\to& \lnu \bbar j \;, \label{eq:tbq} \\

tb             &\to& \lnu \bbar \;, \label{eq:tb} \\

tq             &\to& \lnu bj  \label{eq:tq}

\end{eqnarray}

An overview of the cuts made and their effect on the backgrounds are as
follows (with ATLAS-inspired parameters):

Our primary weapon, of course, is b-tagging (through displaced
vertices) and light quark/gluon
rejection, which we depend on to beat down backgrounds (\ref{eq:wjj}) and
(\ref{eq:tq}). Reconstruction of the Higgs boson will reduce these and
the rest of the backgrounds. This is especially true of background
(\ref{eq:wz}), whose significance at the upper end of the $m_H$ range
considered here will strongly depend on the detector resolution.

To reduce the top backgrounds (\ref{eq:tt}--\ref{eq:tq}), we
essentially veto events with any ``extra'' particles, including jets
or a second lepton. The second neutrino of $t\bar t\to WW\bbar\to
\lnu\lnu\bbar$ is
partially vetoed by a transverse mass cut, to require consistency with the
decay of a $W$ boson. A second $e$ or jet is vetoed in the same way
in the forward region. A second $\mu$ in the forward region cannot be
vetoed, and in fact contributes to the apparent $\not{\!\!E_T}$.

\section{Results}

For detector acceptance and resolution, we have modelled the ATLAS
detector\cite{atlas} on a parton level\footnote{Except that the
constant term in lepton resolution was taken to be twice as large.}.
The  $\not{\!\!\!E_T}$ for each process was constructed assuming a
hermetic calorimeter, and adding the $p_T$ of all jets and leptons
(except for muons beyond the $\mu$-detector) smeared as indicated:
\begin{eqnarray}
\Delta E_{\rm had} / E_{\rm had}   &=&  \frac{50\%}{\sqrt{E_{\rm had}}}
                                            \oplus  3\% \\
\Delta E_{\rm \ell}/E_{\rm \ell}   &=&  \frac{10\%}{\sqrt{E_{\rm \ell}} }
                                            \oplus  2\%

\end{eqnarray}
The detailed cuts are:
\begin{eqnarray}
p_T(l) > 20\,\gev,& |\eta(l)| < 2.5 & \hbox{$e,\mu$
tagging}\;,\label{eq:tag(l)}\\
p_T(b) > 30\,\gev,& |\eta(b)| < 2.0 & \hbox{b-tagging}\;,\label{eq:tag(b)}\\
p_T(\mu) > 10\,\gev,& |\eta(\mu)| < 3.0 & \hbox{extra $\mu$
veto}\;,\label{eq:veto(mu)}\\

p_T(e) > 10\,\gev,& |\eta(e)| < 4.5 & \hbox{extra $e$ veto}\;,
\label{eq:veto(e)}\\

p_T(j) > 10\,\gev,& |\eta(j)| < 4.5 & \hbox{extra jet
veto}\;,\label{eq:veto(j)}\\
|m(\bbar)-m_H| &<\,\, 7.5\,\gev  & \hbox{$m_H$
reconstruction}\;,\label{eq:m(bb)}\\
m_T(\ell,\not{\!p_T}) &<\,\, 80\,\gev & \hbox{$W$-boson consistency cut}
\label{eq:mt}
\end{eqnarray}

The $b$-tagging efficiency is assumed to be $30\%$; the mistagging rate
for light quarks/gluons is taken to be $1\%$.

Total event rates with all the above cuts (for 10 fb$^{-1}$ of
integrated luminosity) are provided in Table~1, for a range of $m_t$ and
$m_H$. The last row gives the number of years at $\int {\cal L} \,dt =
10$ fb$^{-1}$ per year for $5\sigma$ detection. Because of the light
quark/gluon jet rejection, the principal background is process
(\ref{eq:wbb}), while the combined top quark backgrounds account for
roughly half. The signal falls from 67.2 ($m_H = $ 100 GeV) to 40
events ($m_H = 120$ GeV) principally
because of the falling BR($H\to\bbar$). For $m_t = 175$ GeV,
approximately the central value of CDF's search for the top
quark\cite{cdf}, detection at the 5$\sigma$ level would require around
2-4 years for the range of $m_H$ considered here.

As noted above, without a factor of $10^{-4}$ suppression of
background (\ref{eq:wjj}), this process would have swamped the signal.
Doubling the mistagging rate would quadrapule and double processes
(\ref{eq:wjj}) and (\ref{eq:tbq}) respectively. Similarly, if jet
reconstruction has an appreciably lower efficiency around 10 GeV
for the forward compared to the central region, backgrounds
(\ref{eq:tt}--\ref{eq:tbq}) will grow in relative importance.

\section{Improving the S/B ratio}
We note that it is possible, by employing certain cuts, to increase
the S/B ratio at the cost of a small decrease in significance. This
may be a price worth paying, since significance is calculated here
using only statistical uncertainty, and in the end, results may well
be dominated by systematic errors in measuring the $m(\bbar)$
distribution.

Top reconstruction (and vetoing) is an obvious strategy to try, given that
processes (\ref{eq:tt}--\ref{eq:tq}) make up around 1/3 to 1/2 of the
total background. Another strategy is to consider topological cuts.
In figures 1 and 2 are shown two distributions that can be helpful in
increasing the S/B ratio by specifically reducing the top backgrounds.

Backgrounds (\ref{eq:wjj}--\ref{eq:wz}) and backgrounds
(\ref{eq:tt}--\ref{eq:tq}), respectively, fall in the same way for
large $\sum |p_T(b)|$ and small $\Delta R(b,\bar b)$. For example, for
$m_t = 150$ GeV, $m_H = 100$ GeV, a cut of $\Delta R(b,\bar b) < 1.6$ yields
about 17 signal events (for 10 fb$^{-1}$), a total of 30 events from the
first set of backgrounds, and 5 from the top-quark production
background; the S/B ratio is increased from 1/5 to 1/2, while the
significance falls from 3.4 to 2.9. A more complete appraisal of this
and other issues raised above is given elsewhere \cite{us}.

\section{Acknowlegements}

We would like to thank A. Stange and S. Willenbrock for discussion of
their work. This work was supported in part by the National Science Foundation
under Grant No. PHY-9307980

\vspace{0.25 in}

\vskip .1in
\hskip -.35in
\begin{tabular}{||c|r|r|r|r|r|r||} \hline
  &  \multicolumn{2}{c|} {$\mt = 150$ GeV}
  &  \multicolumn{2}{c|} {$\mt = 175$ GeV}
  &  \multicolumn{2}{c||} {$\mt = 200$ GeV} \\  \cline{2-7}
  &  $\mH $   &  $\mH$  &  $\mH$
       &  $\mH $ &  $\mH$ &  $\mH$ \\
  Processes &  100 GeV &  120 GeV &100 GeV
       &  120 GeV &  100 GeV &  120 GeV \\ \hline
$WH$   &  67.2    &  40.0   &  67.2   &  40.0   & 67.2    &  40.0  \\
$Wjj$     & 44.4     &   41.1  &  44.4     &   41.1  & 44.4     &   41.1  \\
$Wbb$     & 113.6 & 82.2   &   113.6 & 82.2   & 113.6  & 82.2   \\
$WZ$      &   45.8 &  1.6  &   45.8  &  1.6 &   45.8 &  1.6 \\
$tt$      &  73.6  &  72.3  &  31.9     &  33.2 &  12.9  & 14.8   \\
$tbq$     &  44.6    &   43.6 & 39.6 &  36.6   &   34.6   &   35.4 \\
$tb$      &  44.8   &  43.4  &  23.6  & 25.8  &   12.6  & 15.0   \\
$tq$      &  22.7  &  24.7  &   16.0   &  19 & 10.3  &   13.3 \\  \hline
  S/B     &   $67/389$   & $40/309$ & $67/315$   & $40/240$
& $67/274$  & 40/203   \\
  Signif. & 3.4 &  2.3  & 3.8 &  2.6 & 4.0  & 2.8  \\   \hline
$\int{\cal L}dt$ (10 fb$^{-1}$) & 2.2 & 4.7 &  1.7 & 3.7 & 1.6 & 3.2 \\

for $5\sigma$ signif. &  &  &    &   &   &  \\ \hline
\end{tabular}
\vspace{.2in}

{\small

Table 1. Event rates (for 10 fb$^{-1}$) at the LHC,
$\sqrt{s} = 14$ TeV with the kinematic cuts specified in the text. The
$b$-tagging efficiency/mistagging rates are taken to be $30\%$ and

$1\%$ respectively.}

\vspace*{2.55in}
{\small

Fig. 1. Differential cross-section $d\sigma/d\sum_{i=b,\bar
b}|p_t(b_i)|$ (pb/5 GeV).}

\newpage
\vspace*{3in}
{\small

Fig. 2. Differential cross-section $d\sigma/d\Delta R(b,\bar b)$ (pb/0.1).}

\vspace*{1 in}
\bibliographystyle{unsrt}

\end{document}